\documentclass{article}

\usepackage{amsmath}
\usepackage[dvips]{geometry}
\usepackage{fullpage}
\usepackage[T1]{fontenc}
\usepackage{graphicx}
\usepackage{SIunits}
\usepackage{url}

\newcommand{\J}{\mathrm{j}}  % square root of -1
\newcommand{\D}{\mathrm{d}}  % differential
\newcommand{\E}{\mathrm{e}}  % exponential

\newcommand{\hgf}[2]{{_{#1}F_{#2}}}
\newcommand{\hankel}[1]{H_{#1}^{(1)}}

\begin{document}

\bibliographystyle{unsrt}

% \title[Ring source]{Series expansion for the sound field of a ring
%   source}
\title{Series expansion for the sound field of a ring source}
\author{Michael Carley} 
% \affiliation{Department of Mechanical
%   Engineering, University of Bath, Bath BA2 7AY, England}
% \email{m.j.carley@bath.ac.uk} \date{\today}

\maketitle

\begin{abstract}
  An exact series expansion for the field radiated by a monopole ring
  source with angular variation in source strength is derived from a
  previously developed expression for the field from a finite
  disk. The derived series can be used throughout the field, via the
  use of a reciprocity relation, and can be readily integrated to find
  the field radiated by arbitrary circular sources of finite extent,
  and differentiated to find the field due to higher order sources
  such as dipoles and quadrupoles.
\end{abstract}
%\pacs{43.28.Ra,43.40.Rj,43.20.Rz}

%\keywords{ring source, rotating systems, Helmholtz equation, Green's
%  function, series expansion, Bessel function, Hankel function}

%\maketitle

\section{Introduction}
\label{sec:intro}

Many problems in acoustics are related to the sound radiated or
scattered by systems with axial symmetry. These systems include rotors
such as cooling fans and aircraft propellers, circular ducts,
vibrating bodies such as baffled speakers, and bodies of rotation. In
each case, to predict the radiated noise, there is a requirement to
compute the field produced by an elemental ring source of given
radius, frequency and angular dependence. This calculation is an
essential part of many noise prediction methods and there is a need
for efficient techniques to perform it.

A second motivation for the study of ring sources is their use as a
model problem for propeller and rotor noise. There are a number of
approximations to the ring source field, developed, in the main, to
examine the nature of the field and its variation with operating
parameters\cite{prentice92}, or to give a far-field approximation for
use in noise prediction\cite{gutin48}. These approximations have
proven useful for industrial noise prediction, such as in aircraft
propeller noise, and form the basis of many practical prediction
techniques.

An approach which does not seem to have found much favour is the use
of exact series expansions for the field of the ring source. There are
numerous such expansions for disc sources of finite extent with
examples covering a number of different
configurations\cite{mast-yu05,carley06a,mellow06,mellow08}. The number
of published expansions for a ring source is quite small, however. One
is the method of Matviyenko\cite{matviyenko95} which gives a five term
recursion for the ring source of a given azimuthal order. A second,
very recent paper, is that of Conway and Cohl\cite{conway-cohl10}
which gives series expansions for the ring source, in terms of Bessel
and Hankel functions and associated Legendre functions. These series
are accurate and easily implemented but they are expressed in terms of
modified variables, of the type used in elliptic integral solutions of
ring potential problems, or of toroidal type. This makes the series
hard to interpret and complicates the issue of differentiating them to
find the field due to higher order sources such as the dipoles and
quadrupoles employed in rotor noise problems. It also makes it awkward
to integrate the series over radius to give an expansion for a finite
disc source.

In this paper, we take a previously published\cite{carley06a}, quite
simple, series for a finite disc source and use it to derive an
expansion for the field from a ring source. The derivation depends on
routine use of mathematical tables and yields an expansion expressed
in physical variables which can, if necessary, be integrated to give a
series for the field of a finite source with arbitrary radial
variation in source strength.

\section{Analysis}
\label{sec:analysis}

\begin{figure}
  \centering
  \includegraphics{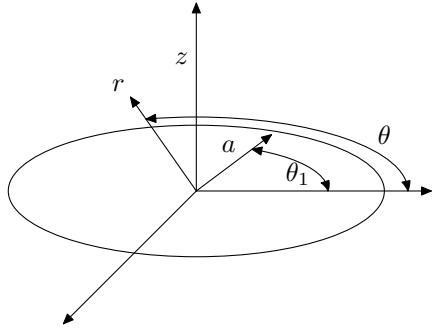}
  \caption{Coordinate system for calculations}
  \label{fig:coordinate}
\end{figure}

The problem to be considered is shown in
Figure~\ref{fig:coordinate}. In cylindrical coordinates
$(r,\theta,z)$, we require the field radiated by a ring monopole
source at radius $a$ in the plane $z=0$, with source strength $\exp[\J
n \theta_{1}]$. The field for wavenumber $k$ is then $\exp[\J
n\theta]R_{n}(k,a,r,z)$:
\begin{align}
  \label{equ:ring}
  R_{n}(k,a,r,z) &= 
  \int_{0}^{2\pi}
  \frac{\E^{\J(kR'+n\theta_{1})}}{4\pi R'}\,\D\theta_{1},\\
  R' &= 
  \left[
    r^{2} + a^{2} - 2ra\cos\theta_{1} + z^{2}
  \right]^{1/2}.\nonumber  
\end{align}
We note that there is a reciprocity relation such that $R_{n}$ is
unchanged if $a$ and $r$ are switched.

A simple, exact series expansion for $R_{n}$ can now be derived, using
previously developed results for a finite disc source. The starting
point is the integral expression for the field
radiated by a disc source $r\leq a$:
\begin{align}
  \label{equ:disc}
  I_{n}(k,a,r,z) &= \int_{0}^{a} 
  R_{n}(k,r_{1},r,z)r_{1}\,
  \D r_{1}.
\end{align}
which has an exact series expansion\cite{carley06a}:
\begin{align}
  I_{n} &= 
  (\pi a^{2})^{1/2} \frac{(-ka)^{n+1/2}}{(kR)^{1/2}} \sum_{m=0}^{\infty}
  A_{m}
  \hankel{n+2m+1/2}(kR)
  P_{n+2m}^{n}(\cos\phi)\nonumber\\
  &\times\hgf{1}{2}
  \left[
    \frac{n+2m+2}{2};
    \frac{n+2m+4}{2},n+2m+3/2;
    -\left(\frac{ka}{2}\right)^{2}
  \right]  
  \left(
    \frac{ka}{2}
  \right)^{2m+1/2}
  \label{equ:expansion}\\
  A_{m} &= (-1)^{m}
  \frac{2^{2m-1}}{n+2m+2}  
  \frac{(2m-1)!!}{(2n+2m)!!(2n+4m-1)!!},\nonumber
\end{align}
where $R=[r^{2}+z^{2}]$ is the distance of the observer from the
origin, $\phi=\tan^{-1}r/z$ is the polar angle of the observer,
$\hankel{\nu}$ is the Hankel function of the first kind of order
$\nu$, $P_{m}^{q}$ is the associated Legendre function,
$\hgf{1}{2}(\cdot)$ is a generalized hypergeometric
function\cite{slater66}:
\begin{align}
  \label{equ:hgf}
  \hgf{1}{2}(a;b,c;x) &= \sum_{n=0}^{\infty} B_{n}x^{n},\\
  B_{n} &= \frac{(a)_{n}}{(b)_{n}(c)_{n}}\frac{1}{n!}\nonumber,
\end{align}
and $(a)_{n}=\Gamma(a+n)/\Gamma(a)$ is Pochammer's
symbol\cite{gradshteyn-ryzhik80}.

Differentiating with respect to $a$ gives an expression for the field
radiated by a ring source of radius $a$:
\begin{align}
  \label{equ:diff}
  \frac{1}{a}\frac{\partial I_{n}}{\partial a} &= R_{n}(k,a,r,z).
\end{align}

Likewise, differentiating Equation~\ref{equ:expansion}:
\begin{align}
  \label{equ:diff:1}
  \frac{\partial I_{n}}{\partial a} &= 
  \J^{2n+1}\frac{\pi^{1/2}}{(kR)^{1/2}}
  \sum_{m=0}^{\infty}
  A_{m}\hankel{n+2m+1/2}(kR)
  P^{n}_{n+2m}(\cos\phi)
  \frac{(ka)^{2m+n+1}}{2^{2m-1/2}}
  \nonumber\\
  &\times\left[
    \frac{n+2m+2}{2}\hgf{1}{2}
    -
    \left(\frac{ka}{2}\right)^{2}
    \hgf{1}{2}'
  \right],
\end{align}
where the parameters and argument of the hypergeometric function have
been suppressed for compactness and the derivative $\hgf{1}{2}'$ is
taken with respect to the argument. Using the relation:
\begin{align*}
  \hgf{1}{2}(a_{1};b_{1},b_{2};x)a_{1} +
  \hgf{1}{2}'(a_{1};b_{1},b_{2};x)x &= \hgf{1}{2}(a_{1}+1; b_{1},
  b_{2}; x)a_{1},
\end{align*}
Equation~\ref{equ:diff:1} can be rewritten:
\begin{align*}
  \frac{\partial I_{n}}{\partial a} &= 
  \J^{2n+1}\frac{\pi^{1/2}}{(kR)^{1/2}}
  \sum_{m=0}^{\infty}
  A_{m}\hankel{n+2m+1/2}(kR)
  P^{n}_{n+2m}(\cos\phi)
  \frac{(ka)^{2m+n+1}}{2^{2m+1/2}}
  \nonumber\\
  &\times
  (n+2m+2)\hgf{1}{2}
  \left[
    \frac{n+2m+4}{2};
    \frac{n+2m+4}{2},n+2m+3/2;
    -    \left(\frac{ka}{2}\right)^{2}
  \right],
\end{align*}
noting the change in the first parameter of the hypergeometric
function.

Using the cancellation property of hypergeometric functions:
\begin{align*}
  \hgf{1}{2}(a_{1}; a_{1}, b_{2}; x) &= \hgf{0}{1}(;b_{2}; x)
\end{align*}
and the relation\cite{watson95}:
\begin{align*}
  \hgf{0}{1}\left[
    ; \nu+1; -\left(\frac{z}{2}\right)^{2}
  \right] &= 
  \Gamma(\nu+1)\left(\frac{2}{z}\right)^{\nu}J_{\nu}(z),
\end{align*}
where $J_{\nu}$ is a Bessel function of the first kind,
\begin{align}
  \hgf{1}{2}
  \left[
    \frac{n+2m+4}{2};
    \frac{n+2m+4}{2},n+2m+3/2;
    -\left(\frac{ka}{2}\right)^{2}
  \right]
  &= \nonumber\\
  \label{equ:hgf:bessel}
  \pi^{1/2}\frac{(2n+4m+1)!!}{2^{n+2m+1}}
  \left(
    \frac{2}{ka}
  \right)^{n+2m+1/2}
  J_{n+2m+1/2}(ka),
\end{align}
we find an expansion for the field radiated by a ring source:
\begin{align}
  R_{n}(k,a) = \frac{1}{a}\frac{\partial I_{n}}{\partial a} = 
  \J^{2n+1}\frac{\pi}{4}
  \frac{1}{(aR)^{1/2}}
  &\sum_{m=0}^{\infty}
  (-1)^m
  \frac{(2n+4m+1)(2m-1)!!}{(2n+2m)!!}\nonumber\\
  &\times
  \label{equ:result}
  \hankel{n+2m+1/2}(kR)  P_{n+2m}^{n}(\cos\phi)
  J_{n+2m+1/2}(ka)  
\end{align}
Equation~\ref{equ:result} is the main result of the paper. It is an
exact series expansion for the field radiated by an oscillating ring
source of radius $a$ to any point with $R\geq a$. If it is required to
compute the field at points $R<a$, this can be done using the
reciprocity relation which allows switching of $r$ and $a$ in the ring
source integral of Equation~\ref{equ:diff}. The expansion is
remarkably simple, containing only one special function,
$P_{n+2m}^{n}$, given that the Bessel and Hankel functions can be
evaluated as finite sums of elementary
functions\cite{gradshteyn-ryzhik80}:
\begin{align*}
  J_{n+1/2}(x) &= 
  \left(\frac{2}{\pi x}\right)^{1/2}
  \sum_{q=0}^{n}
  \frac{1}{q!}
  \frac{(n+q)!}{(n-q)!}\frac{\cos[x-(n-q+1)\pi/2]}{(2x)^{q}},\\
  \hankel{n-1/2}(x) &= 
  \J^{-n}
  \left(
    \frac{2}{\pi x}
  \right)^{1/2}
  \E^{\J x}
  \sum_{q=0}^{n-1} \frac{(-1)^{q}}{q!} \frac{(n+q-1)!}{(n-q-1)!}
  \frac{1}{(\J 2 x)^{q}}.
\end{align*}
We further note that this expansion has a form very similar to that of
the expansion of the Helmholtz Green's function for a point source
using the ``summation theorem'' for Bessel
functions\cite{gradshteyn-ryzhik80}.
%8.533

\section{Results and applications}
\label{sec:results}

\begin{figure*}
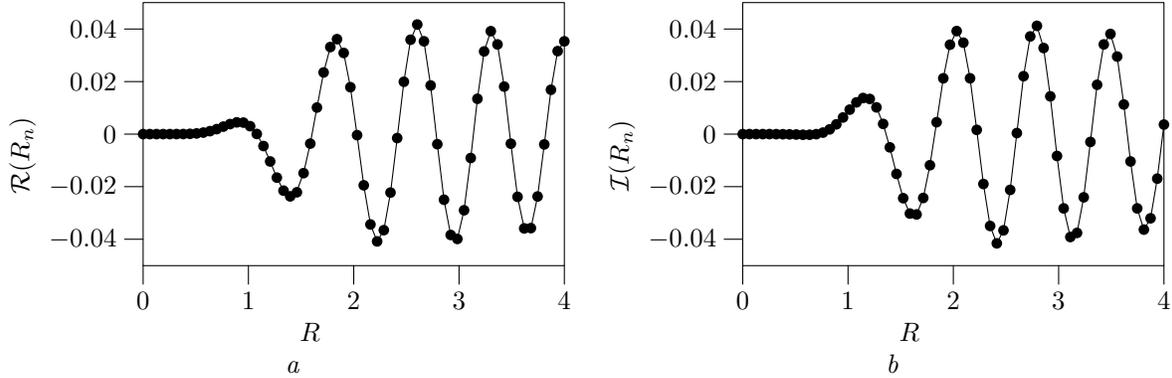

  \centering
  \begin{tabular}{cc}
    \includegraphics{jasa10-figs.3} &
    \includegraphics{jasa10-figs.4} \\
    \textit{a} & \textit{b}
  \end{tabular}
  \caption{Comparison of numerical (solid) and series (circle)
    computation of $R_{n}$, $\phi=\pi/6$; \textit{a}: real part;
    \textit{b}: imaginary part.}
  \label{fig:numerical:30}
\end{figure*}

\begin{figure*}
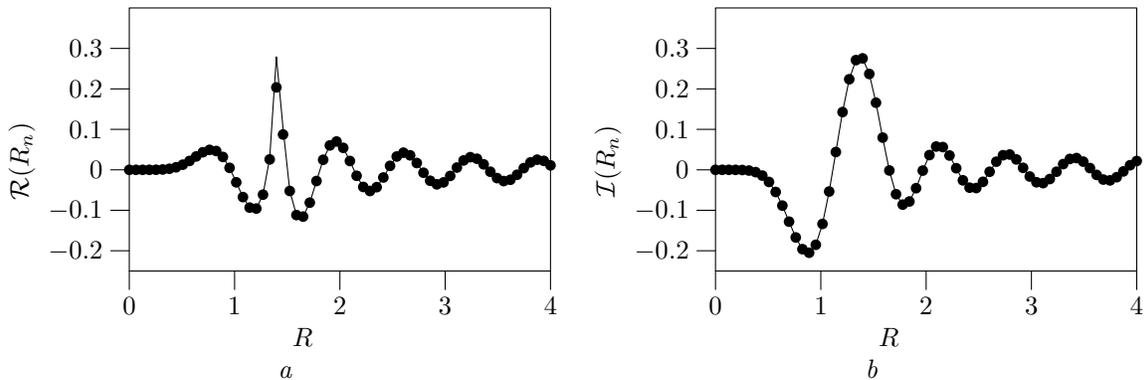

  \centering
  \begin{tabular}{cc}
    \includegraphics{jasa10-figs.5} &
    \includegraphics{jasa10-figs.6} \\
    \textit{a} & \textit{b}
  \end{tabular}
  \caption{Comparison of numerical (solid) and series (circle)
    computation of $R_{n}$, $\phi=\pi/2$; \textit{a}: real part;
    \textit{b}: imaginary part.}
  \label{fig:numerical:90}
\end{figure*}

As a check on the accuracy of Equation~\ref{equ:result}, some
calculations were performed for arbitrary values of the parameters and
compared to direct numerical evaluation of
$R_{n}$. Figure~\ref{fig:numerical:30} shows the real and imaginary
parts of $R_{n}$ computed numerically and using the series expansion
for $k=10$, $n=7$, $a=2^{1/2}$, along a ray $0\leq R\leq 4$,
$\phi=\pi/6$. For points $R<a$, the series was evaluated employing the
reciprocity relation and switching $r=R\sin\phi$ and $a$. As is clear
from the plots, the two results are identical and the series is seen
to be an accurate method of evaluating the
field. Figure~\ref{fig:numerical:90} shows data for the same case but
in the source plane, $\phi=\pi/2$. Again, the results are practically
identical to those from numerical evaluation. It is worth noting that
the singularity near the source radius $a=2^{1/2}$ has been accurately
captured. In the discussion of their series expansion, Conway and
Cohl\cite{conway-cohl10} note that one of their series, which uses
Hankel functions, is accurate far from the ring but not close to it,
whereas their Legendre function series is accurate near the ring,
since the Legendre function has a built-in singularity there, but is
slow to converge in the far field. Equation~\ref{equ:result}, being
based on both Hankel and Legendre functions, is able to capture the
behaviour in both regions and is accurate and rapidly convergent over
the full range considered. 

\subsection{Finite disc source}
\label{sec:finite}

A major application of a ring source evaluation method is in rotor
acoustics where the radiated field is given by integrals of the form:
\begin{align}
  \label{equ:rotor}
  p_{n}(k,a,r,z) &= \int_{0}^{a} 
  s(r_{1})R_{n}(k,r_{1},r,z)r_{1}\, \D r_{1},
\end{align}
where $s(r_{1})$ is a radial source function whose value depends on
the rotor geometry and/or loading.

For points lying outside the sphere containing a rotor of radius $a$,
substitution of Equation~\ref{equ:result} into
Equation~\ref{equ:rotor} gives a series expansion for the acoustic
field of the rotor:
\begin{align}
  p_{n} =
  \J^{2n+1}\frac{\pi}{4}
  \frac{1}{R^{1/2}}
  &\sum_{m=0}^{\infty}
  (-1)^m
  \frac{(2n+4m+1)(2m-1)!!}{(2n+2m)!!}\nonumber\\
  &\times
  \label{equ:rotor:series}
  \hankel{n+2m+1/2}(kR)  P_{n+2m}^{n}(\cos\phi)s_{n+2m},\\
%  J_{n+2m+1/2}(ka)  
  s_{n+2m} &= \int_{0}^{a} s(r_{1})
  J_{n+2m+1/2}(kr_{1})r_{1}^{1/2}\,\D r_{1},\nonumber
\end{align}
so that the coefficients $s_{n+2m}$ are given by a Hankel transform of
the radial source term. It has been known for many years that the
far-field noise from a rotor is given by a Hankel transform of the
radial source term\cite{gutin48,wright69}, but using a single Bessel
function of integer order rather than those of order integer plus one
half used in Equation~\ref{equ:rotor:series} which is exact at all
points in the field $R>a$, including the near field.

% \begin{itemize}
% \item Expression for unit source field.
% \end{itemize}

% \subsection{Higher order sources}
% \label{sec:higher}

% \begin{itemize}
% \item dipole by differentiation.
% \end{itemize}

\section{Conclusions}
\label{sec:conclusions}

A simple exact series expansion for the acoustic field radiated by a
monopole ring source has been developed, derived from a previous
result for a finite disc. The series has been tested numerically and
compared to another recently published expansion for the Green's
function for a Helmholtz problem in cylindrical coordinates. Since it
is based on physical variables, the series is easily integrated to
give an expansion for finite sources, such as rotors, with arbitrary
radial variation in source strength, and is also easily differentiated
to find the fields due to higher order sources such as dipoles and
quadrupoles.

\bibliography{abbrev,propnoise,misc,maths,scattering}

\begin{thebibliography}{10}
\newcommand{\enquote}[1]{``#1''}
\expandafter\ifx\csname url\endcsname\relax
  \def\url#1{\texttt{#1}}\fi
\expandafter\ifx\csname urlprefix\endcsname\relax\def\urlprefix{URL }\fi
\providecommand{\bibinfo}[2]{#2}
\providecommand{\noopsort}[1]{}
\providecommand{\switchargs}[2]{#2#1}

\bibitem{prentice92}
\bibinfo{author}{P.~R. Prentice}, \enquote{\bibinfo{title}{The acoustic ring
  source and its application to propeller acoustics}},
  \bibinfo{journal}{Proceedings of the Royal Society of London. \textrm{A}.}
  \textbf{\bibinfo{volume}{437}}, \bibinfo{pages}{629--644}
  (\bibinfo{year}{1992}).

\bibitem{gutin48}
\bibinfo{author}{L.~Gutin}, \enquote{\bibinfo{title}{On the sound field of a
  rotating propeller}}, \bibinfo{type}{Technical Memorandum}
  \bibinfo{number}{1195}, \bibinfo{institution}{NACA},
  \bibinfo{address}{Langley Aeronautical Laboratory, Langley Field, Va. USA}
  (\bibinfo{year}{1948}).

\bibitem{mast-yu05}
\bibinfo{author}{T.~D. Mast} and \bibinfo{author}{F.~Yu},
  \enquote{\bibinfo{title}{Simplified expansions for radiation from a baffled
  circular piston}}, \bibinfo{journal}{Journal of the Acoustical Society of
  America} \textbf{\bibinfo{volume}{118}}, \bibinfo{pages}{3457--3464}
  (\bibinfo{year}{2005}).

\bibitem{carley06a}
\bibinfo{author}{M.~Carley}, \enquote{\bibinfo{title}{Series expansion for the
  sound field of rotating sources}}, \bibinfo{journal}{Journal of the
  Acoustical Society of America} \textbf{\bibinfo{volume}{120}},
  \bibinfo{pages}{1252--1256} (\bibinfo{year}{2006}).

\bibitem{mellow06}
\bibinfo{author}{T.~Mellow}, \enquote{\bibinfo{title}{On the sound field of a
  resilient disk in an infinite baffle}}, \bibinfo{journal}{Journal of the
  Acoustical Society of America} \textbf{\bibinfo{volume}{120}},
  \bibinfo{pages}{90--101} (\bibinfo{year}{2006}).

\bibitem{mellow08}
\bibinfo{author}{T.~Mellow}, \enquote{\bibinfo{title}{On the sound field of a
  resilient disk in free space}}, \bibinfo{journal}{Journal of the Acoustical
  Society of America} \textbf{\bibinfo{volume}{123}},
  \bibinfo{pages}{1880--1891} (\bibinfo{year}{2008}).

\bibitem{matviyenko95}
\bibinfo{author}{G.~Matviyenko}, \enquote{\bibinfo{title}{On the azimuthal
  {Fourier} components of the {Green's} function for the {Helmholtz} equation
  in three dimensions}}, \bibinfo{journal}{Journal of Mathematical Physics}
  \textbf{\bibinfo{volume}{36}}, \bibinfo{pages}{5159--5169}
  (\bibinfo{year}{1995}).

\bibitem{conway-cohl10}
\bibinfo{author}{J.~T. Conway} and \bibinfo{author}{H.~S. Cohl},
  \enquote{\bibinfo{title}{Exact {Fourier} expansion in cylindrical coordinates
  for the three-dimensional {Helmholtz} {Green} function}},
  \bibinfo{journal}{Zeitschrift f{\"{u}}r angewandte Mathematik und Physik}
  (\bibinfo{year}{2010, in press, available from:
  \url{http://dx.doi.org/10.1007/s00033-009-0039-6}, last viewed 15 April
  2010}).

\bibitem{slater66}
\bibinfo{author}{L.~J. Slater}, \emph{\bibinfo{title}{Generalized
  hypergeometric functions}} (\bibinfo{publisher}{Cambridge University Press},
  \bibinfo{address}{Cambridge}) (\bibinfo{year}{1966}).

\bibitem{gradshteyn-ryzhik80}
\bibinfo{author}{I.~Gradshteyn} and \bibinfo{author}{I.~M. Ryzhik},
  \emph{\bibinfo{title}{Table of integrals, series and products}},
  \bibinfo{edition}{5th} edition (\bibinfo{publisher}{Academic},
  \bibinfo{address}{London}) (\bibinfo{year}{1980}).

\bibitem{watson95}
\bibinfo{author}{G.~N. Watson}, \emph{\bibinfo{title}{A treatise on the theory
  of {Bessel} functions}} (\bibinfo{publisher}{Cambridge University Press},
  \bibinfo{address}{Cambridge}) (\bibinfo{year}{1995}).

\bibitem{wright69}
\bibinfo{author}{S.~E. Wright}, \enquote{\bibinfo{title}{Sound radiation from a
  lifting rotor generated by asymmetric disk loading}},
  \bibinfo{journal}{Journal of Sound and Vibration}
  \textbf{\bibinfo{volume}{9}}, \bibinfo{pages}{223--240}
  (\bibinfo{year}{1969}).

\end{thebibliography}

\end{document}